\documentclass[%
reprint,
superscriptaddress,
amsmath,amssymb,
aps,
prb,
longbibliography,
]{revtex4-1}

\usepackage{pdfpages} 
\usepackage{dirtytalk}
\usepackage{dsfont} 
\usepackage{mathtools}
\usepackage{pdfpages}
\usepackage{mhchem}
\usepackage{graphicx}
\usepackage{dcolumn}
\usepackage{bm}

\usepackage{hyperref}
\hypersetup{
colorlinks = true,   
urlcolor = blue,    
linkcolor = magenta,   
citecolor = blue  
}
\usepackage{float}
\usepackage{color}

\makeatletter
\AtBeginDocument{\let\LS@rot\@undefined}
\makeatother

\begin{document}
 \title{Corner- and sublattice-sensitive Majorana zero modes on the kagome lattice}
 \author{Majid Kheirkhah}
 \affiliation{Department of Physics, University of Alberta, Edmonton, Alberta T6G 2E1, Canada}
 \author{Di Zhu}
 \affiliation{School of Physics, Sun Yat-Sen University, Guangzhou 510275, China}
 \author{Joseph Maciejko}
 \affiliation{Department of Physics, University of Alberta, Edmonton, Alberta T6G 2E1, Canada}
 \affiliation{Theoretical Physics Institute, University of Alberta, Edmonton, Alberta T6G 2E1, Canada}
 \author{Zhongbo Yan}
 \email{yanzhb5@mail.sysu.edu.cn}
 \affiliation{School of Physics, Sun Yat-Sen University, Guangzhou 510275, China}
\date{\today}
\begin{abstract}
In a first-order topological phase with sublattice degrees of freedom, a change in the boundary sublattice termination has no effect on the existence of gapless boundary states in dimensions higher
than one. However, such a change may strongly affect the physical properties of those boundary states.  Motivated by this observation, we perform a systematic study of the impact of sublattice terminations on
the boundary physics on the two-dimensional kagome lattice. We find that the energies of the Dirac points of helical edge states in two-dimensional first-order topological kagome insulators sensitively depend on the terminating sublattices at the edge. Remarkably, this property admits the realization of a time-reversal invariant 
second-order topological superconducting phase with highly controllable Majorana Kramers pairs at the corners and 
sublattice domain walls by putting the topological kagome insulator in proximity to
a $d$-wave superconductor. Moreover, substituting the $d$-wave superconductor with a conventional $s$-wave superconductor, we find that 
highly controllable Majorana zero modes can also be realized at the corners and sublattice domain walls if 
an in-plane Zeeman field is additionally applied. Our study reveals promising platforms to implement highly
controllable Majorana zero modes.  
\end{abstract}

\pacs{Valid PACS appear here}
\maketitle
\section{Introduction}
Over the past two decades, topological insulators (TIs) and topological superconductors (TSCs)
have attracted a great amount of interest owing to their novel properties and
promising applications in a diversity of fields~\cite{hasan2010,qi2011topological,leijnse2012introduction,stanescu2013majorana,sato2016majorana,Haim2019}. The most salient property of
TIs and TSCs is the so-called bulk-boundary correspondence,
which describes the one-to-one correspondence between the number of robust gapless states on the boundary of the sample
and the topological invariant characterizing the bulk energy bands~\cite{Chiu2015RMP}.
Very recently, TIs and TSCs have been further refined due to  the discovery of robust gapless boundary states with higher codimension $d_{c}$ in
some insulators and superconductors originally thought to be topologically trivial.  Concretely, insulators and superconductors whose gapless boundary states have $d_{c}=n$
are classified as $n$th-order TIs~\cite{Benalcazar2017,Schindler2018,Benalcazar2017prb,Song2017,Langbehn2017hosc,Zhang2013surface,Khalaf2018,Geier2018,ezawa2018higher,Franca2018,Dumitru2019,Wang2019hoti,Xu2019hoti,Park2019hoti,Liu2019hoti,Sheng2019hoti,Chen2020hoti,PhysRevB.104.184403} or TSCs~\cite{Zhu2018hosc,Yan2018hosc,Wang2018hosc,Wangyuxuan2018hosc,Liu2018hosc,Shapourian2018SOTSC,Hsu2018hosc,Wuzhigang2019hosc,Yan2019hosca,Volpez2019SOTSC,Zhu2019mixed,Peng2019hinge,Ghorashi2019,Zhang2019hinge,Zhang2019hoscb,Yan2019hoscb,Bultinck2019,Franca2019SOTSC,Pan2019SOTSC,Majid2020hosca,Majid2020hoscb,Wu2020SOTSC,Hsu2019HOSC,Wu2020BOTSC,Laubscher2020hosc,Tiwari2020,Ahn2020hosc,Bitan2020hosc,Li2021BTSC,Niu2020hosc,Wuxianxin2021hosc,Fu2021hinge,Luo2021hosc,Jahin2021,Qin2021SOTSC,Tan2021hosc,You2019,Zhang2020mzm,Zhang2020tqc,Pahomi2020braiding,Bomantara2020tqc,Ikegaya2021,Majid2021vortex,Ghosh2021hosc,Li2021hosc,roy2021mixed,Majid2022hosc,wu2022majorana,Ghosh2021fsotc}. Since conventional TIs and TSCs
have $d_{c}=1$,
they are also dubbed first-order topological phases.
For $d_{c}=2$, the bulk-boundary correspondence is also dubbed as bulk-corner correspondence
in two dimensions and bulk-hinge correspondence in three dimensions. This is because the robust gapless
states are expected to be bound at some sharp corners in two dimensions and some
hinges in three dimensions. In experiments, such bulk-corner correspondence and bulk-hinge correspondence
have been widely taken as a smoking gun to verify the realization of
higher-order topological phases~\cite{Peterson2018,Serra-Garcia2018,Imhof2018corner,Schindler2018bismuth,Gray2019helical}.

Besides spin-orbit coupling, it is known that the lattice structure is another important factor
in the formation of first-order topological band structures. For lattices with sublattice degrees of freedom,
topological band structures in general emerge easier since the multiple bands
enforced by the sublattice degrees of freedom commonly support gapless Dirac points~\cite{Asano2011}, the precursor of various first-order
topological gapped phases~\cite{shen2013topological}, if the lattice symmetry groups have higher-dimensional irreducible
representations in the Brillouin zone. The most celebrated example is the honeycomb lattice,
which naturally supports Dirac points in the absence of spin-orbit coupling,
and supports a first-order TI when the Dirac points are gapped by
the presence of spin-orbit coupling~\cite{kane2005a,kane2005b}. While the importance of lattice structure
with sublattice degrees of freedom for the formation of first-order topological band structure is well appreciated,
the impact of sublattices on the gapless boundary states
has attracted much less attention except in one dimension, where the Su-Schrieffer-Heeger (SSH)
model provides an explicit example that the boundary sublattice terminations directly determine the
presence or absence of boundary bound states~\cite{su1979}. This is easy to appreciate, since in dimensions higher
than one, a change of the boundary sublattice termination is known to have no effect on the existence of gapless states in a first-order
topological phase.

Recently, Zhu {\it et al.} showed that while the boundary sublattice terminations in a
two-dimensional (2D) TI with honeycomb lattice do
not affect the existence of helical edge states, they have a dramatic impact on the locations of
boundary Dirac points, which are
the crossing points of the energy dispersion of the helical edge states in the reduced boundary
Brillouin zone~\cite{Zhu2021sublattice}. Remarkably,
they found that if one puts the TI in proximity to an unconventional
superconductor, the sensitive sublattice dependence of boundary Dirac points allows
the realization of sublattice-sensitive Majorana Kramers pairs with codimension $d_{c}=2$
even for a cylindrical-geometry system which has no sharp corners~\cite{Zhu2021sublattice}. This finding
reveals that the sublattice degrees of freedom can have an intriguing interplay
with the second-order topology.

In two dimensions, the kagome lattice is another kind of lattice structure with sublattice degrees of
freedom. In recent years, kagome-lattice materials have attracted great interest as the special lattice
geometry allows a diversity of novel band-structure properties, such as Dirac points,
flat bands, and van Hove singularities~\cite{Li2018kagome,Kang2020kagome,Kang2020b,Li2021kagome}, which allow many interaction-driven competing orders~\cite{Balents2010,Tang2011,Wang2013kagome,Mazin2014}.
Similar to the honeycomb lattice, the kagome lattice
structure also supports topological insulating phases in the presence of strong spin-orbit coupling~\cite{Guo2009}.
Since the honeycomb lattice has only two sublattices
whereas the kagome lattice has three, it is natural to expect
that, with the increase of one more sublattice degree of freedom, the interplay between sublattices 
and second-order topology can result in even richer boundary physics. 
Motivated by this observation, in this work 
we perform a systematic study of the impact of sublattice terminations on
the boundary physics in the kagome lattice.

Starting with a tight-binding model describing the kagome lattice with intrinsic spin-orbit coupling
that realizes a first-order TI~\cite{Guo2009}, we find that, similar to the honeycomb lattice, the sublattice terminations do not 
affect the existence of helical edge states but have a dramatic impact on the boundary Dirac points~\cite{Zhu2021sublattice}. 
Putting the topological kagome insulator in proximity to
a $d$-wave superconductor, the helical edge states are gapped as expected. 
Because of the strong impact of sublattice terminations on the dispersion of 
the helical edge states, we find that the positions of Dirac-mass domain walls binding Majorana 
Kramers pairs sensitively depend on the details of the boundary geometry. When all 
edges are uniform (only having one type of sublattice termination), 
the locations of Majorana Kramers pairs follow the standard picture. 
That is, they appear only at the corners of the system~\cite{Zhu2018hosc,Yan2018hosc,Wang2018hosc,Wangyuxuan2018hosc,Liu2018hosc}. However, when 
two types of sublattice terminations appear on the same edge, we find 
that the sublattice domain walls, corresponding to the interface of two
distinct types of sublattice terminations, can also bind Majorana Kramers pairs. 
Most remarkably, we find that when the sublattice domain walls bind Majorana Kramers pairs, the Majorana Kramers pairs can be tuned to any position on the boundary by controlling the sublattice terminations and they are no longer pinned at certain sharp corners. 
Furthermore, by substituting the $d$-wave superconductor with a conventional $s$-wave superconductor, we find that
highly controllable sublattice-sensitive Majorana zero modes (MZMs) can also be realized if
an in-plane Zeeman field is additionally applied. Our findings reveal that 
the interplay between sublattices and second-order topology in the kagome lattice
can also result in different boundary physics, suggesting the generality of the discussed physics.

This paper is organized as follows. In Sec.~\ref{sec_2}, the kagome lattice and the associated Bogoliubov-de Gennes (BdG) Hamiltonian in the presence of both $d$-wave and conventional on-site $s$-wave pairing are introduced briefly. In Sec.~\ref{sec_3},
we show that the energies of the boundary Dirac points of the topological kagome insulator
 have a sensitive dependence on the sublattice terminations. In Sec.~\ref{sec_4}, it is shown that corner and sublattice-sensitive Majorana Kramers pairs can be realized when the topological kagome insulator is in proximity
to a $d$-wave superconductor. In Sec.~\ref{sec_5}, we illustrate that corner and sublattice-sensitive MZMs can be realized by breaking time-reversal (TR) symmetry via an in-plane Zeeman field
when the topological kagome insulator is in proximity to a conventional $s$-wave superconductor. 
Finally, we conclude with a discussion in Sec.~\ref{sec_6}.
\section{General theoretical formalism}
\label{sec_2}
We consider the growth of a 2D topological kagome insulator on 
a  superconductor and assume that the TI
inherits the pairing symmetry from the bulk superconductor. 
Taking into account the proximity-induced superconductivity, the 
essential topological physics  in the 
2D superconducting topological kagome insulator
can be effectively described by the following minimal BdG Hamiltonian
\begin{eqnarray}
\mathcal{H}_{\rm BdG}(\mathbf{k})=\left(
              \begin{array}{cc}
                \mathcal{H}_{\rm N}(\mathbf{k})-\mu & \Delta(\mathbf{k}) \\
                \Delta^{\dag}(\mathbf{k}) & \mu-\mathcal{H}_{\rm N}^{*}(-\mathbf{k}) \\
              \end{array}
            \right).
\end{eqnarray}
Here, the basis is $\Psi_{\bm{k}}^{\dag}=(\bm{c}^{\dagger}_{\bm{k},\uparrow}, 
\bm{c}^{\dagger}_{\bm{k},\downarrow}, \bm{c}_{-\bm{k},\uparrow}, 
\bm{c}_{-\bm{k},\downarrow})$ with $\bm{c}^{\dagger}_{\bm{k},\sigma} = (
c^{\dagger}_{\bm{k},A,\sigma}, c^{\dagger}_{\bm{k},B,\sigma}, c^{\dagger}_{\bm{k},C,\sigma}
)$ where $c^{\dagger}_{\bm{k},\alpha,\sigma} (c_{\bm{k},\alpha,\sigma})$ denotes a fermionic creation (annihilation) operator
with the subscripts $\alpha=(A, B, C)$ and $\sigma=(\uparrow,\downarrow)$ 
referring to the three sublattices in the unit cell of the 
kagome lattice and  the two spin degrees of freedom, respectively. In the 
BdG Hamiltonian, $\mathcal{H}_{\rm N}(\bm{k})$ refers to the normal-state Hamiltonian, 
$\mu$ is the chemical potential, and $\Delta(\bm{k})$ is the pairing matrix. 
Here, we consider~\cite{Guo2009} 
\begin{eqnarray}
\mathcal{H}_{\rm N}(\mathbf{k})=s_{0}\otimes\mathcal{H}_{\rm 0}(\bm{k})+s_{z}\otimes \mathcal{H}_{\rm soc}(\bm{k})
+s_{x}\otimes \mathcal{H}_{\rm M}
\end{eqnarray}
with 
\begin{align}
\mathcal{H}_{\rm 0}(\bm{k})& =-2t
\begin{pmatrix}
0 && \cos k_1  && \cos k_2 \\
\cos k_1 && 0 && \cos k_3 \\
\cos k_2 && \cos k_3 && 0
\end{pmatrix}, \nonumber
\\
\mathcal{H}_{\rm soc}(\bm{k})&=
2i \lambda
\begin{pmatrix}
0 & \cos(k_2 + k_3) & -\cos(k_1-k_3)\\
& 0 & \cos(k_1 + k_2) \\
&  & 0
\end{pmatrix} + {\rm H.c.},\nonumber
\\
\mathcal{H}_{\rm M}&=
h_x \mathds{1}_{3 \times 3},
\end{align}
where $s_{x,z}$ refer to the Pauli matrices  in spin space, $\mathds{1}_{3 \times 3}$ refers to the $3\times3$ identity matrix 
in sublattice space, $k_i = \bm{k}\cdot \bm{a}_i$ for $i = 1,2,3$ 
where $\bm{a}_1 = \hat{x}$, $\bm{a}_2 = (\hat{x} + \sqrt{3} \hat{y})/2$, 
$\bm{a}_3 = \bm{a}_2 - \bm{a}_1 $, the abbreviation H.c. stands for Hermitian conjugation, and the lattice constants are set to unity for 
notational simplicity. In the normal-state Hamiltonian, $\mathcal{H}_{\rm 0}(\bm{k})$ describes the
kinetic energy originating from nearest-neighbor hoppings, $\mathcal{H}_{\rm soc}(\bm{k})$ 
describes the spin-orbit coupling involving next-nearest-neighbor hoppings, and $\mathcal{H}_{\rm M}$ 
describes the Zeeman splitting induced by an in-plane magnetic field, where the field direction is taken to be along the $x$ direction for the sake of simplicity as the discussed physics turns out 
to be independent of the field direction as long as it is parallel to the system plane.

For the superconducting pairing order parameters, in this work we will focus on 
conventional $s$-wave pairing and $d$-wave pairings which are common in real materials. 
For the $s$-wave pairing, we consider the simplest on-site form, which 
is described by $H_{\rm s}=\sum_{i}\Delta_{0}c_{i,\uparrow}^{\dag}c_{i,\downarrow}^{\dag}+\mathrm{h.c.}$ in real space where
the sublattice degrees of freedom have been made implicit in the real-space coordinates $i=(i_{x},i_{y})$. 
For the $d$-wave pairing, we also consider the simplest nearest-neighbor form, which 
is described by $H_{\rm d}=\frac{1}{2}\sum_{\langle i,j\rangle}\Delta_d \cos(2 \theta_{ij})[c^{\dagger}_{i,\uparrow}
 c^{\dagger}_{j,\downarrow} - c^{\dagger}_{i,\downarrow} c^{\dagger}_{j,\uparrow}]+\mathrm{h.c.}$ in real space, 
 where the $\theta_{ij}$ is the angle between the line connecting two nearest-neighbor sites $i$ and $j$ of the kagome lattice and the positive direction of the $x$ axis. This corresponds to $d_{x^2-y^2}$ pairing, belonging to the 2D $E_2$ irreducible representation of the $C_{6v}$ point group of the kagome lattice. As illustrated in Fig.~\ref{sch_0}, $\theta_{ij} = 0$ between sublattices $A$ and $B$ and vice versa, $\theta_{ij} = \pi/3$ between sublattices $A$ and $C$ and vice versa, and $\theta_{ij} = 2\pi/3$ between sublattices $B$ and $C$ and vice versa.  In momentum space, the considered $s$-wave pairing and $d$-wave pairing can be compactly written in the pairing matrix, which gives
 \begin{widetext}
\begin{equation}
\Delta(\bm{k}) = i s_y \otimes
\begin{pmatrix}
\Delta_0 && 2\Delta_d \cos{k_1} && -\Delta_d \cos{k_2}\\
2\Delta_d \cos{k_1} && \Delta_0 && -\Delta_d \cos{k_3} \\
-\Delta_d \cos{k_2} && -\Delta_d \cos{k_3} && \Delta_0
\end{pmatrix},
\end{equation}
 \end{widetext}
where $\Delta_0$ and $\Delta_d$ represent the on-site $s$-wave and $d$-wave superconducting pairing amplitudes, respectively, and are taken to be real numbers.

\section{Dependence of helical edge states on boundary sublattice terminations in the normal state }
\label{sec_3}

Without $\mathcal{H}_{\rm M}$,
the tight-binding Hamiltonian $\mathcal{H}_{\rm N}(\bm{k})$ corresponds to a generalization of the
Kane-Mele model from the honeycomb lattice to the kagome lattice and describes
a TI as long as
$\lambda\neq 0$~\cite{Guo2009}. As a first-order topological phase, the bulk-boundary correspondence dictates
the existence of a pair of gapless helical states on the open boundary, irrespective of the orientation
or microscopic details of the boundary. This, however, does not mean that the physical properties
of the helical edge states are insensitive to the microscopic details of the boundary.
For the Kane-Mele model on the honeycomb lattice with two sublattices, it was shown that  
the boundary Dirac points would shift
from one TR invariant momentum to the other when the edge changes its termination from one sublattice to
the other~\cite{Zhu2021sublattice}. In addition, the boundary Dirac points would stay at the same energy if there is no potential offset between the two types of sublattices to break the particle-hole symmetry~\cite{Zhu2021sublattice}. 
In comparison, the kagome lattice has three types of sublattices and there is 
no particle-hole symmetry; thus the impact of sublattice terminations may be quite different from the honeycomb lattice.

\begin{figure}[t!]
\centering
\includegraphics[scale= 0.31]{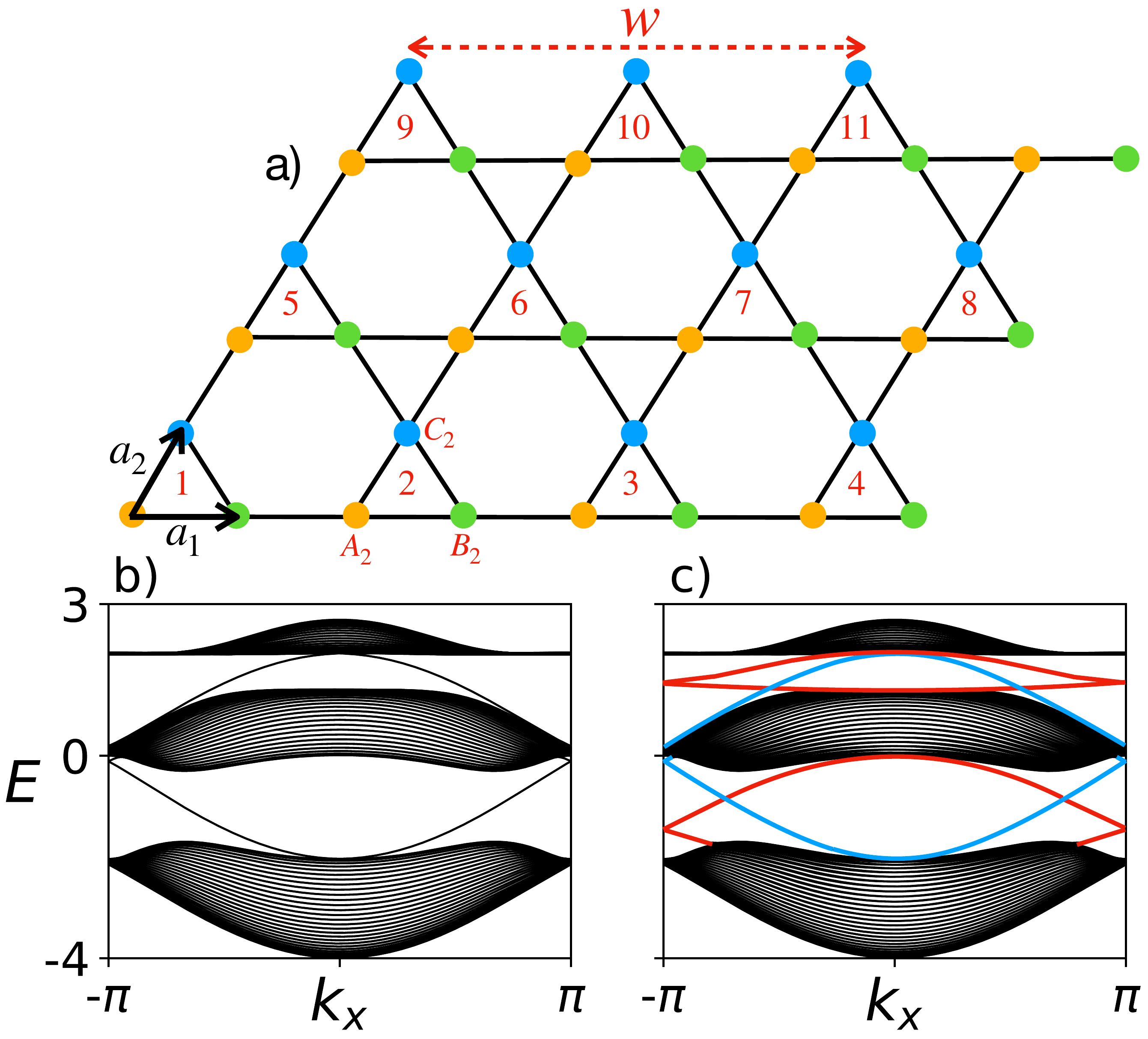}
\caption{(Color online) (a) Schematic figure of different sublattice terminations at the edges of the kagome lattice for $N_x = 4$, $N_y = 3$, and $w = 3$. With open boundary conditions (see Sec.~\ref{sec_4}), the left, bottom, and right edges of the lattice respectively terminate with sublattices $A$ (orange dots) and $C$ (blue dots), $A$ and $B$ (green dots), and $B$. With periodic boundary conditions in the $x$ direction, sublattice terminations refer to the top and bottom edges only. We vary the sublattice termination of the top edge by tuning $w$, i.e., it can be terminated with sublattice $C$ ($A$ and $B$) for $w=N_x$ ($w=0$). A domain wall between the $C$ and $AB$ terminations occurs for $0 < w <N_x$. The triangular unit vectors are denoted by $\boldsymbol{a}_1 = \hat{x}$ and $\boldsymbol{a}_2 = (\hat{x} + \sqrt{3} \hat{y})/2$. (b),(c) Energy spectrum of the normal-state Hamiltonian for a cylindrical geometry with open (periodic) boundary conditions in the $y$ ($x$) direction. Chosen parameters are $t = 1$, $\lambda = 0.2$, $h_x = 0$, $w = 0$ in (b) and $w = N_x$ in (c). In (c), the top edge dispersion is plotted in red and the bottom edge dispersion in blue. The edge states tangentially merge with the bulk states in the energy spectrum.}
\label{sch_0}
\end{figure}

To reveal the impact of sublattice terminations on the helical edge states in the kagome lattice, 
 we diagonalize the Hamiltonian under different boundary conditions.
Without loss of generality, we consider a cylindrical geometry with periodic boundary conditions in the $\boldsymbol{a}_1$ direction and open boundary conditions in the $\boldsymbol{a}_2$ direction; for simplicity, we refer to those directions as the $x$ and $y$ directions, respectively. For this geometry, it is easy to see from Fig.~\ref{sch_0}(a) that a flat $y$-normal edge has two types of sublattice terminations; one being the $AB$ type ($w = 0$) and the other being the $C$ type ($w = N_x$). Here $0\leq w\leq N_x$ denotes the number of $C$ sites on the edge.

When both the top and bottom $y$-normal edges have the $AB$-type termination,
the numerical results show that the energy dispersion of the helical edge states
on the top and bottom edges are degenerate, as shown in Fig.~\ref{sch_0}(b), which is a result of the inversion symmetry for such a choice of boundary conditions. Upon
modifying the termination of the top edge from $AB$ type to $C$ type,
the numerical results show that the degeneracy is strongly lifted as a result of inversion symmetry breaking, as shown in
Fig.~\ref{sch_0}(c). Comparing Fig.~\ref{sch_0}(c) with Fig.~\ref{sch_0}(b),  it is readily found that the boundary Dirac points for the helical boundary states on the top edge (red dispersions) have a considerable energy shift when the termination is changed from $AB$ type to $C$ type, suggesting
that the properties of the helical edge states also have a sensitive dependence
on the boundary termination in the kagome lattice.

Before proceeding, it would be illuminating to 
make a further comparison of the boundary physics in the honeycomb and kagome lattices. 
For the honeycomb lattice, each edge only supports one boundary Dirac point since 
the helical edge states only exist in a window of the reduced boundary Brillouin zone~\cite{kane2005a,kane2005b}. 
With the change of sublattice terminations, the boundary Dirac point will shift 
from one TR invariant momentum to the other in 
the reduced boundary Brillouin zone. This shift of boundary Dirac point can be intuitively understood by noting that if one views the boundary momentum as a parameter, the 2D 
particle-hole symmetric Kane-Mele Hamiltonian can be viewed as a parameter-dependent one-dimensional Hamiltonian, and it corresponds to a spinful SSH model with opposite topological properties (or say opposite dimerization) at the two TR invariant boundary momenta~\cite{su1979,asboth2016short,Zhu2021sublattice}. For the kagome lattice, a close inspection of Figs.~\ref{sch_0}(b) and \ref{sch_0}(c) reveals that the helical edge states exist in 
the whole reduced boundary Brillouin zone. Furthermore, 
the boundary Dirac points for the $AB$-type edge (blue dispersions) overlap with the bulk states in energy, 
while the boundary Dirac points for the $C$-type edge (red dispersions) are well located in the energy gap. 
The large energy shift in the boundary Dirac points from one sublattice termination to the other
is due to the absence of any symmetry constraint on the energy of boundary Dirac points.
For a given edge, the energy offset between Fermi level and boundary Dirac point directly affects the locations
of boundary Fermi points. Since the properties of the Fermi surface play a crucial role in a weak-coupling 
superconducting system, it is natural to expect that the sublattice-sensitive large energy shift of boundary Dirac points 
also allows the realization of sublattice-sensitive Majorana bound states in a topological kagome insulator that is proximity-coupled to a superconductor.

\section{Majorana Kramers pairs at corners and sublattice domain walls}
\label{sec_4}
MZMs at the corners of an open system are viewed as a defining
property of second-order TSCs in two dimensions~\cite{Zhu2018hosc,Yan2018hosc,Wang2018hosc,Wangyuxuan2018hosc,Liu2018hosc}. In previous works, it was shown that,
when a TI without sublattice degrees of freedom is in proximity
to a $d$-wave superconductor, Majorana Kramers pairs will emerge at the system's corners
if the edges forming the corners are along crystallographic directions that are separated by a line node of the $d$-wave pairing function in momentum space~\cite{Yan2018hosc}. The underlying physics can be simply understood via the
Jackiw-Rebbi theory~\cite{Jackiw1976}. That is, the emergence of Majorana Kramers pairs is a consequence of 
the formation of Dirac-mass domain walls at the corners. More precisely, $d$-wave superconductivity 
will introduce a Dirac mass to gap out the helical edge states, and the sign of the Dirac mass will follow the 
angular dependence of the $d$-wave pairing~\cite{Yan2018hosc}. If the line-node direction of the $d$-wave pairing
is along the $(11)$ and $(1\bar{1})$ directions, then a square lattice with edges in the 
$(01)$ and $(10)$ directions will carry four Majorana Kramers pairs, one per each of the four corners~\cite{Yan2018hosc}. Since the pairing line-node direction is fixed, if the helical 
edge states are not sensitive to local deformations of the boundary, then 
the Dirac-mass domain walls in such systems, and thus the locations of the Majorana Kramers pairs, 
will also  not be sensitive to such changes.

Above, we have shown that the sublattice terminations have a strong impact on the dispersion of helical edge states in topological kagome insulators. A natural question 
is whether those sublattice terminations also have a strong impact on corner MZMs. 
To compare with known results, we first focus on the TR invariant case ($h_{x}=0$) and 
also consider $d$-wave pairing.  

\begin{figure}[t!]
\centering
\includegraphics[scale= 0.525]{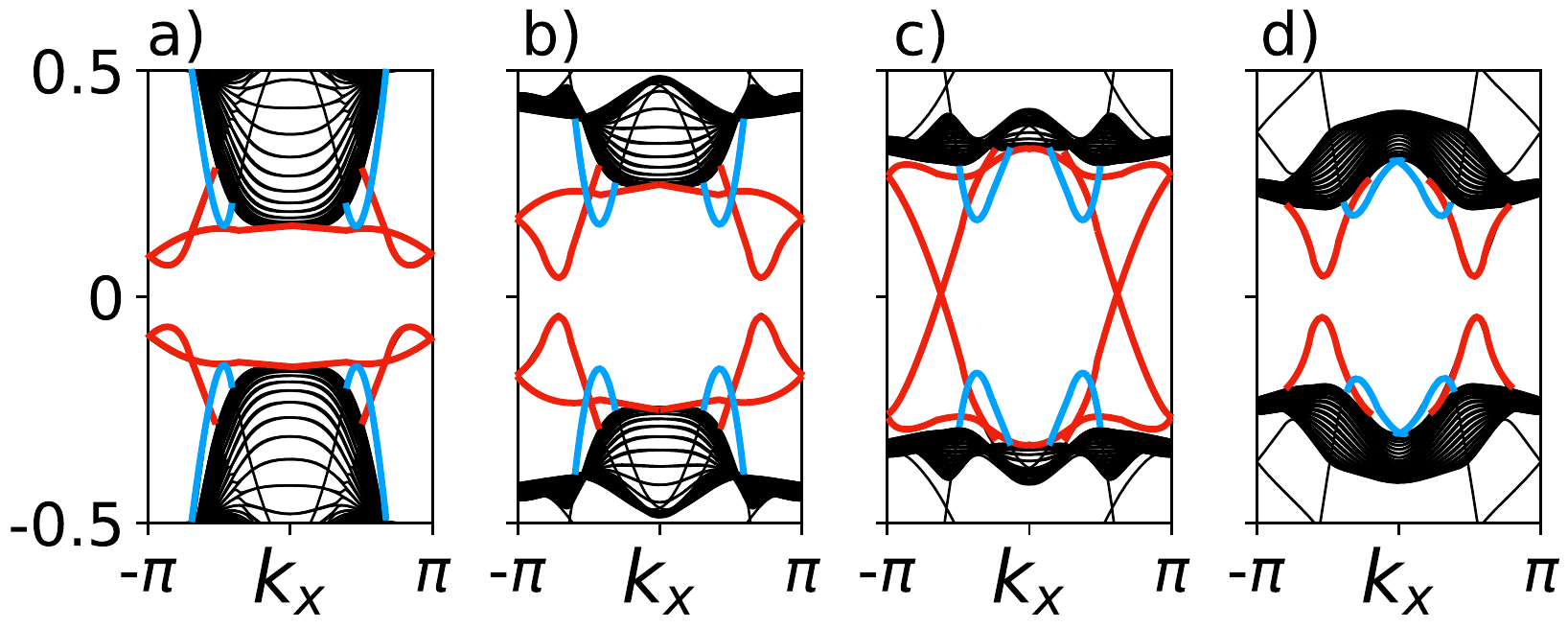}
\caption{(Color online) Energy spectrum of the BdG Hamiltonian with $d$-wave superconducting pairing for a cylindrical geometry with open (periodic) boundary conditions in the $y$ ($x$) direction, and the bottom (top) edge terminates with sublattices $A$ and $B$ ($C$). The top edge dispersion is plotted in red and the bottom edge dispersion in blue. The chosen parameters are $t = 1$, $\lambda = 0.2$, $\Delta_d = 0.1$, and $\mu = 1.5$ in (a), $\mu = 1.6$ in (b), $\mu = 1.7$ in (c), and $\mu = 1.8$ in (d). Tuning $\mu$ closes and reopens the boundary energy gap, signaling a boundary topological phase transition at $\mu = 1.7$.}
\label{fig_2}
\end{figure}

As before, we first consider a cylindrical geometry with periodic boundary conditions in the $x$ direction
and open boundary conditions in the $y$ direction. To see the difference 
for edges with different terminations, we consider the two $y$-normal edges to be different, 
with the top one being the $C$ type and the lower one being the $AB$ type. The normal-state band structure is then given in Fig.~\ref{sch_0}(c). In Fig.~\ref{fig_2}, we
choose the chemical potential to be located within the gap between the top band and the middle band
and show the evolution of the BdG energy spectra with respect to the chemical potential. 
In Fig.~\ref{fig_2}(a), we choose the
chemical potential to be located near the upper Dirac point at $k_{x}=\pi$  shown in Fig.~\ref{sch_0}(c). 
As expected, the numerical result shows that $d$-wave pairing opens a gap in both the top (red) and bottom (blue) edge dispersions. The resulting gaps, however, have different magnitudes on the top and bottom
edges. With the increase of chemical potential (tuning the chemical potential further away 
from the boundary Dirac point at $k_{x}=\pi$), we find that the top edge will undergo a closing-and-reopening 
transition in the energy gap, while the energy gap for the bottom edge remains finite in the whole range of chemical potentials considered, as shown in Figs.~\ref{fig_2}(b)-\ref{fig_2}(d).

The distinct evolution of the boundary BdG energy spectrum with respect 
to the chemical potential on the top $C$-type edge and bottom $AB$-type edge 
can be understood via a low-energy edge theory. At a qualitative level, 
the low-energy physics on the edges is described by 
\begin{eqnarray}
H_{C/AB}(q_{\alpha})&=&\epsilon_{\alpha}(q_{\alpha})\tau_{z}\otimes s_{0}+ v_{\alpha}q_{\alpha}\tau_{z}\otimes s_{z}\nonumber\\
&&-\delta\mu_{\alpha}\tau_{z}\otimes s_{0}+\Delta_{\rm eff,\alpha}(q_{\alpha})\tau_{y}\otimes s_{y},
\end{eqnarray}
where $q_{\alpha}$ denotes the momentum measured from the TR
invariant momentum $k_{x,\alpha}=0/\pi$, $\epsilon_{\alpha}$ describes the particle-hole asymmetry of the Dirac cone at $k_{x,\alpha}$, 
$v_{\alpha}$ refers to the corresponding velocity, $\delta\mu_{\alpha}$ denotes the chemical potential 
measured from the corresponding Dirac point, and $\Delta_{\rm eff,\alpha}$ describes the boundary pairing, which is determined by 
projecting the bulk pairing onto the subspace spanned by the wave functions of helical edge states. 
Based on the low-energy Hamiltonian, the boundary energy gap closes 
when the boundary pairing nodes determined by $\Delta_{\rm eff,\alpha}(q_{\alpha})=0$ coincide with the boundary Fermi 
points determined by $\pm v_{\alpha}q_{\alpha}=\delta\mu_{\alpha}-\epsilon_{\alpha}(q_{\alpha})$. 
Without needing to know the detailed forms of $\epsilon_{\alpha}$, $v_{\alpha}$, $\delta\mu_{\alpha}$, and $\Delta_{\rm eff,\alpha}$, 
one can infer from Fig.~\ref{sch_0}(c) that the critical Fermi momenta at which the boundary energy gap closes should be different on the two types of edges. For the range of chemical potential considered in Fig.~\ref{fig_2}, 
the change of Fermi momenta on the $C$-type edge covers a considerable fraction of the Brillouin zone, 
while the change of Fermi momenta on the $AB$-type edge only covers a small range.
This explains why the boundary energy gap undergoes a closing-and-reopening transition on the $C$-type edge
but remains open on the $AB$-type edge when the chemical potential varies in the range considered.

It is worth noting that the closing-and-reopening transition suggests a change in the boundary topology on the top edge. Since a 1D
TR invariant superconductor follows a $\mathbb{Z}_{2}$ classification~\cite{schnyder2008,kitaev2009periodic}, if the top and bottom 
edges have the same topology before the closing-and-reopening transition, then they can have different topology after 
the transition and vice versa. When the top and bottom edges have different topology, the system will 
inevitably harbor Dirac-mass domain walls on the boundary if the remaining direction is also cut open. 

\begin{figure}[t!]
\centering
\includegraphics[scale= 0.42]{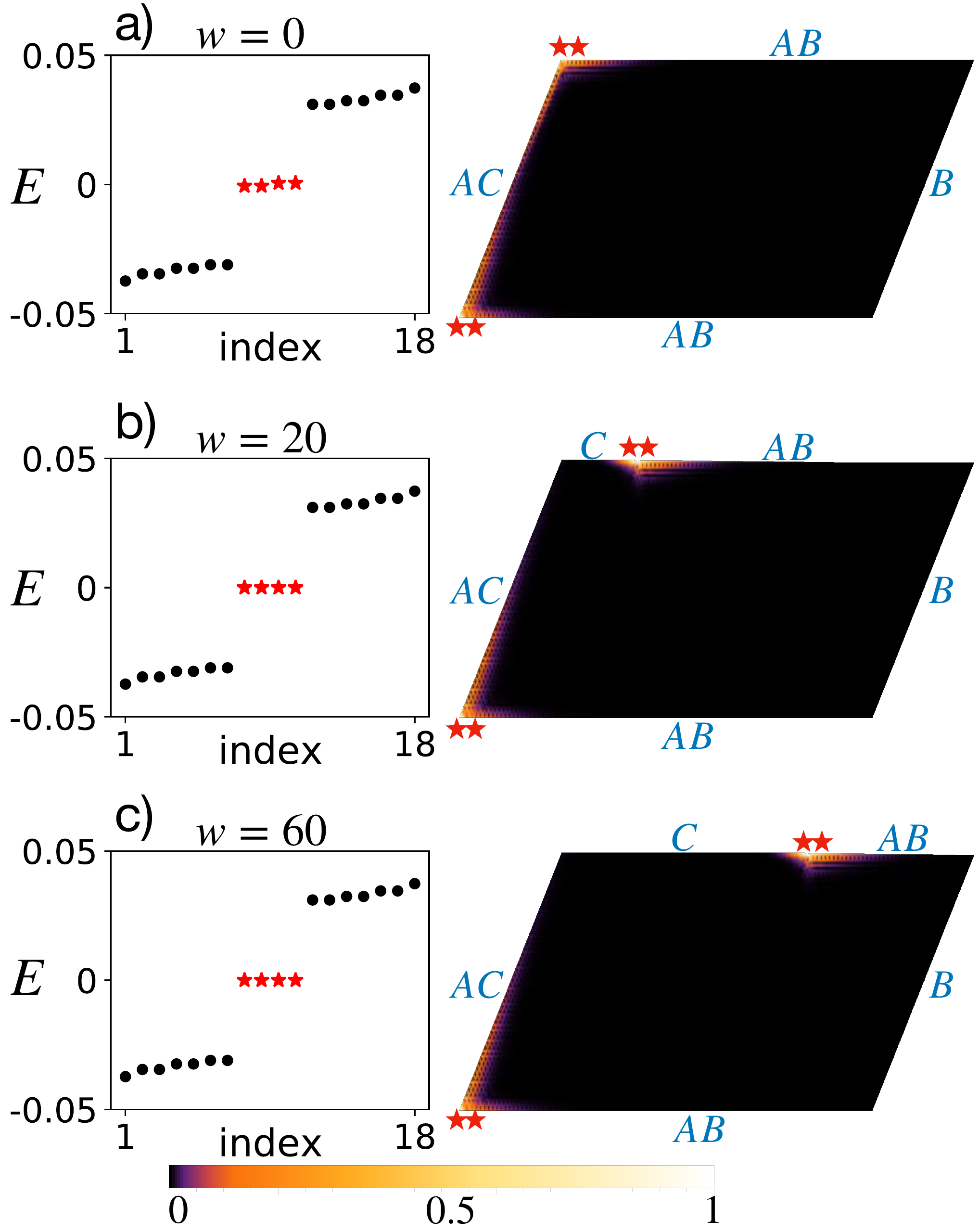}
\caption{(Color online) Majorana Kramers pairs at the corners and sublattice domain walls of the kagome lattice, for top-edge sublattice terminations with (a) $w=0$, (b) $w=20$, and (c) $w=60$ [see Fig.~\ref{sch_0}(a)]. The two (four) red stars in the left panels represent one (two) Majorana Kramers pair; their corresponding probability densities and positions are plotted in the right panels. In (b) and (c), there is one Majorana Kramers pair at the $C$-$AB$ sublattice domain wall. Chosen parameters are $t = 1$, $\lambda = 0.2$, $\mu = 1.442$, $\Delta_d = 0.1$, $h_x = 0$, $N_x = 100$, and $N_y = 50$. The positions of the Majorana Kramers pairs can be precisely manipulated by locally controlling the sublattice terminations.}
\label{fig_3}
\end{figure}

Our results for the cylindrical geometry confirm that the sublattice terminations have a strong impact
on the boundary BdG energy spectrum and thus the boundary topology of the superconducting state. Next we consider a geometry with open boundary
conditions in both $x$ and $y$ directions and investigate the impact of sublattice terminations on the locations
of Majorana Kramers pairs. Without loss of generality, we choose the left boundary  to be an $AC$-type edge
and the right boundary to be a $B$-type edge, as shown in Fig.~\ref{sch_0}(a). We choose the chemical potential to cross the upper boundary Dirac point at $k_{x}=\pi$, in the gap between the two uppermost bands [Fig.~\ref{sch_0}(c)].
When both the top and bottom $y$-normal boundaries are $AB$-type edges,
the numerical result shows that there are only two Majorana Kramers pairs. Furthermore, they are localized at two corners corresponding to the ends of the left $AC$-type edge, as
shown in Fig.~\ref{fig_3}(a). The result in Fig.~\ref{fig_3}(a) suggests that the Dirac mass on the left $AC$-type edge has a sign opposite to the other three edges. It is worth noting 
that the properties of helical edge states on the $AC$-type edge and $AB$-type edge 
are the same since the two edges are related by a $120^{\circ}$ rotation and the normal Hamiltonian 
is invariant under such a rotation. Thus the sign difference 
between $AC$-type edge and $AB$-type edge can be intuitively understood through the angular 
dependence of $d$-wave pairing, whereas the sign difference between $AC$-type edge and $B$-type edge
needs to be understood via the change of boundary topology induced by the change of 
sublattice terminations illustrated in Fig.~\ref{fig_2}. Because of the strong impact of sublattice terminations, remarkably,
we find that the locations of the Majorana Kramers pairs will move with the local change of sublattice
terminations, as shown in Fig.~\ref{fig_3}(b) and \ref{fig_3}(c).

The displacement of Majorana Kramers pairs with the local change of sublattice termination can be simply understood from the fact that this change can modify the boundary topology of a given edge, as illustrated in Fig.~\ref{fig_2}. Within a low-energy description,
it is known that the boundary physics can be effectively described by a real-space linear Dirac Hamiltonian of the form \cite{shen2013topological,Yan2018hosc}
\begin{equation}
\mathcal{H}=-\frac{i}{2}\Gamma_{1}\{v(l),\partial_l\}+M(l)\Gamma_{2},
\end{equation}
where $v(l)$ denotes the velocity whose
value depends on the boundary coordinate $l$ ($l$ is defined modulo the boundary perimeter)
but the sign is fixed, $M(l)$ denotes the Dirac mass, and $\Gamma_{1,2}$ are $4\times 4$ anticommuting Dirac matrices. Note that, because $v(l)$ is space dependent, the first term of this equation is symmetrized to make it Hermitian. It is worth emphasizing that this symmetrized form, just like the standard form~\cite{shen2013topological}, supports solutions for Majorana bound states at the Dirac-mass domain walls where $M(l)$ changes sign.
A change in the local boundary topology corresponds to a
local change of the sign of the Dirac mass $M(l)$, thus leading to
a change in the positions of Dirac-mass domain walls binding Majorana Kramers pairs.

The results in Fig.~\ref{fig_3} reveal an important property of second-order topological systems
with sublattice degrees of freedom: bound states with codimension $d_{c}=2$ can be manipulated to any position on the boundary due to the sensitive sublattice dependence.
This is quite different from systems without sublattice degrees of freedom, where the bound states are necessarily pinned at corners.
 
\section{Majorana zero modes at corners and sublattice domain walls}
\label{sec_5}

When TR symmetry is broken by an in-plane Zeeman field, second-order topological superconductivity can be realized even with a conventional $s$-wave superconductor in proximity 
to the TI~\cite{Wu2020SOTSC}.  Without sublattice degrees 
of freedom, previous works showed that the resulting MZMs will again be pinned
at the system corners if the edges of the system are sharp~\cite{Wu2020SOTSC}. In this section, 
we are going to investigate whether the presence of sublattice degrees of freedom will modify 
this picture. 

\begin{figure}[t!]
\centering
\includegraphics[scale= 0.525]{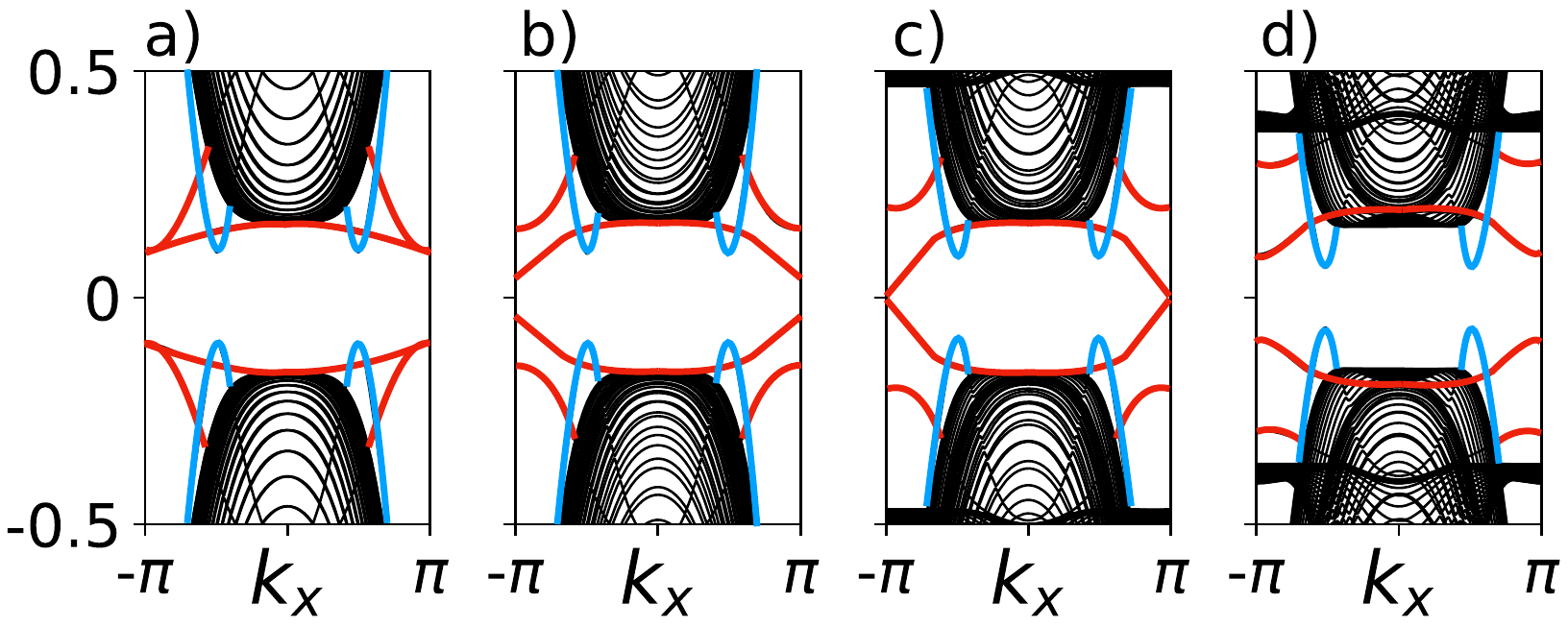}
\caption{(Color online) Energy spectrum of the BdG Hamiltonian with on-site $s$-wave superconducting pairing for a cylindrical geometry with open (periodic) boundary conditions in the $y$ ($x$) direction, and the bottom (top) edge terminates with sublattices $A$ and $B$ ($C$). The top edge dispersion is plotted in red and the bottom edge dispersion in blue. Chosen parameters are $t = 1$, $\lambda = 0.2$, and $\mu = 1.442$, $\Delta_0 = 0.1$, and $h_x = 0$ in (a), $h_x = 0.05$ in (b), $h_x = 0.1$ in (c), and $h_x = 0.2$ in (d). On-site $s$-wave pairing opens a gap in the edge dispersion and increasing the Zeeman field $h_x$ closes and reopens this gap, signaling a boundary topological phase transition at $h_x = 0.1$.}
\label{fig_4}
\end{figure}

In parallel to the TR symmetric case, we first consider
a similar cylindrical geometry with periodic boundary conditions in the $x$ direction
and open boundary conditions in the $y$ direction, and the top and bottom $y$-normal edges
are also respectively set to the $C$ and $AB$ types. Without loss of generality, we choose
the Fermi level to cross the boundary Dirac point of the top edge, in the uppermost band gap, and investigate the evolution
of the BdG energy spectra with respect to the Zeeman field. For a fixed pairing strength, we find
that $s$-wave pairing also opens a gap in the helical edge states on both the top and the bottom
edges, as shown in Fig.~\ref{fig_4}(a). With the increase of Zeeman field, we find that the energy gap on the top edge undergoes a closing-and-reopening transition at the TR invariant momentum $k_{x}=\pi$,
while the energy gap on the bottom edge stays open in the whole range of field considered, as shown in Figs.~\ref{fig_4}(b)-\ref{fig_4}(d).
The critical value of the Zeeman field at the transition follows a formula similar to the widely known form 
in superconducting nanowire systems~\cite{lutchyn2010,oreg2010helical}, i.e., $h_{c}=\sqrt{(\delta\mu)^{2}+\Delta_{0}^{2}}$, where
$\delta\mu$ refers to the energy offset between the Fermi level and the boundary Dirac point.
For $s$-wave pairing, because the Dirac mass induced by pairing is uniform on the boundary,
there will be no Dirac-mass domain walls binding Majorana Kramers pairs
on the boundary when the Zeeman field is absent. In the presence of Zeeman field, 
since a 1D superconductor without TR
symmetry also follows a $\mathbb{Z}_{2}$ classification~\cite{schnyder2008,kitaev2009periodic}, Dirac-mass domain walls binding MZMs
will inevitably emerge when the Zeeman field drives some of the edges into a different topological regime.

\begin{figure}[t!]
\centering
\includegraphics[scale= 0.415]{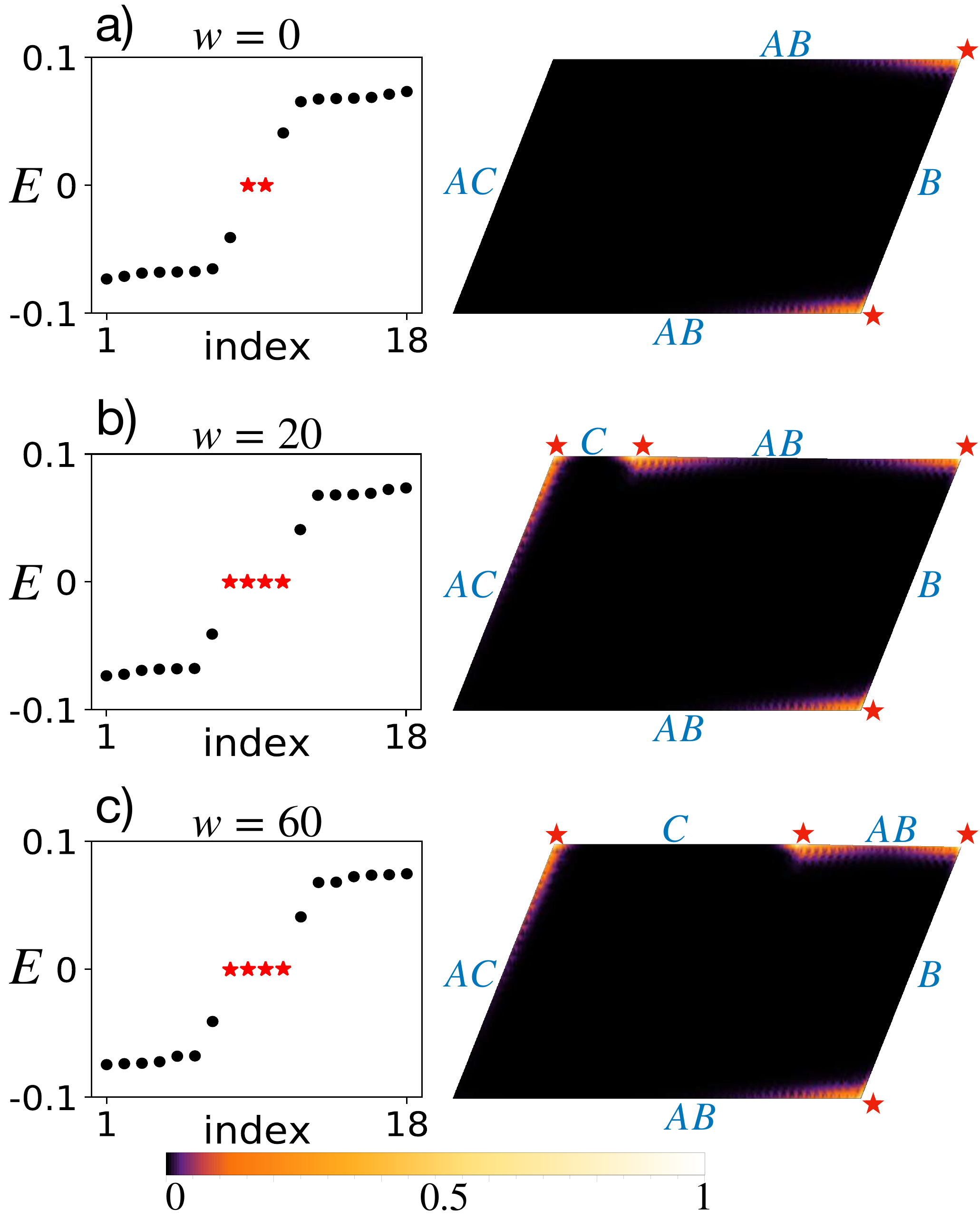}
\caption{(Color online) MZMs at the corners and sublattice domain walls of the kagome lattice, for top-edge sublattice terminations with (a) $w=0$, (b) $w=20$, and (c) $w=60$ [see Fig.~\ref{sch_0}(a)]. The red stars in the left panels represent MZMs; their corresponding probability densities and positions are plotted in the right panels. In (b) and (c), there is one MZM at the $C$-$AB$ sublattice domain wall. Chosen parameters are $t = 1$, $\lambda = 0.2$, $\mu = 1.442$, $\Delta_0 = 0.1$, $h_x = 0.2$, $N_x = 100$, and $N_y = 50$. The positions of the MZMs can be precisely manipulated by locally controlling the sublattice terminations.}
\label{fig_5}
\end{figure}

To investigate the impact of sublattice termination on the MZMs for this 
TR symmetry breaking case, we next 
consider a geometry with open boundary conditions in both $x$ and $y$ directions. 
Without loss of generality, similar to the $d$-wave pairing case, we consider that the left and right boundaries are fixed to be $AC$-type and $B$-type edges, respectively, and the bottom $y$-normal edge is fixed to the $AB$ type. 
In Fig.~\ref{fig_5}(a), the numerical results show that, when the top $y$-normal edge also assumes the $AB$-type 
termination, there are only two MZMs with their wave functions mainly localized at the 
two corners corresponding to the two ends of the right $B$-type edge. The locations of the MZMs can be simply understood by noting that the $AB$-type edge and $AC$-type edge should
have the same topology. This can be inferred from the following facts. 
First, as aforementioned, the $AB$-type edge and $AC$-type edge are related by a $60^{\circ}$ rotation. 
Second, we have previously pointed out that the physical properties of the concerned
system are invariant under the rotation of the in-plane Zeeman field, 
so the physical properties of the BdG Hamiltonian with on-site $s$-wave pairing
are invariant under such a rotation. On the other hand, we have shown in Fig.~\ref{fig_4} 
that the Zeeman field will induce a closing-and-reopening transition in 
the boundary energy gap on the $C$-type edge. Since the $B$-type edge is
also related to the $C$-type edge by a $60^{\circ}$ rotation, their boundary 
topology should be the same. Thus, the boundary topology on the $B$-type edge is 
distinct from the other three edges for the set of parameters considered, leading to the formation of MZMs 
at the two corners shown in Fig.~\ref{fig_5}(a).

By changing part of the top $y$-normal edge's sublattice 
termination to the $C$ type from the left, we find the emergence of a new pair of MZMs,
as shown in Fig.~\ref{fig_5}(b). 
Furthermore, the MZM located at 
the sublattice domain walls between $C$-type and $AB$-type terminations 
is found to track the location of the sublattice domain walls, as shown in Fig.~\ref{fig_5}(c). The additional presence and movement of the MZMs can also be simply understood by  noting that the boundary topology on the $C$-type edge is distinct to the $AB$-type and $AC$-type edge. It is worth noting that, in Fig.~\ref{fig_5}, 
the wave functions of MZMs display a much stronger localization on the $B$-type and $C$-type edges than on the $AB$-type and $AC$-type edges. The remarkable difference in localization on different edges is a result of the remarkable difference in boundary energy spectra [see Fig.~\ref{fig_4}(c)]. Roughly speaking, the MZM wave functions obey the approximate Jackiw-Rebbi form,
\begin{align}
\psi(x)\varpropto e^{-\int^{x}\frac{M(x')}{v}dx'}.
\end{align}
Thus their localization properties are controlled by the ratio of Dirac mass $M$ to velocity $v$. Along $C$ (or $B$) edges, the boundary gap (and thus Dirac mass) is larger and the velocity is smaller than along the $AB$ (or $AC$) edges [Fig.~\ref{fig_4}(d)]. Therefore, the MZMs are more strongly localized along the $C$ or $B$ edges than along the $AB$ or $AC$ edges.

The emergence of additional MZMs and their displacement 
with the local change of sublattice terminations are also 
due to the sensitive dependence of the boundary topology on the sublattice terminations. The 
numerical results above reveal that,
for this TR symmetry breaking case, 
all MZMs can also in principle be tuned away from the corners to any position on the boundary due to their sensitive dependence on sublattice terminations.

\section{Discussions and Conclusions}
\label{sec_6}
Bound states pinned at corners of an open system have been widely 
taken as a characteristic  property of second-order topology in two dimensions. 
For systems without sublattice degrees of freedom, this scenario generally holds, 
because the boundary topology is not sensitive to the local 
change of boundary geometry. Rather, the change of boundary topology associated with the formation of Dirac-mass domain walls can only occur at certain corners of the system. In this work, we considered heterostructures composed 
of first-order TIs with kagome lattice structure and 
superconductors in proximity, and investigated the impact of boundary sublattice 
terminations on the locations of Majorana bound states. Because of the three-sublattice structure of the kagome lattice, we found that the boundary 
topology has a sensitive dependence on sublattice terminations. Such a sensitive sublattice dependence allows Majorana bound states to be positioned anywhere on the boundary, rather than being pinned at the corners
of the system. This remarkable tunability can benefit the detection of Majorana 
bound states as well as their manipulation
for applications to topological quantum computing.

Our findings for the kagome lattice can be straightforwardly generalized to 
the 2D Lieb lattice, owing to their many similarities in physical 
properties~\cite{Weeks2010a,Jiang2019kagome}.  Together with a previous study on the honeycomb lattice~\cite{Zhu2021sublattice}, 
our present study reveals that sublattice degrees of freedom 
have a strong and interesting interplay with second-order topology. 
Since sublattice degrees of freedom are ubiquitous in real materials, 
the existence of sublattice-sensitive bound states  
is expected to be a common property of systems with sublattice degrees
of freedom and second-order topology. In the past few years, 
a considerable amount of kagome materials have been found to exhibit 
nontrivial topological properties~\cite{Ye2018,Yin2020,Ortiz2020}. 
With the fast accumulation of material candidates, 
our predictions are expected to be testable in the near future. In experiments, we suggest the use of scanning tunneling microscopy or scanning force 
microscopy to manipulate the sublattice terminations~\cite{Stroscio1991,Custance2009}.
\begin{acknowledgments}
M.Kh. and J.M. acknowledge support from NSERC Discovery Grant No.~RGPIN-2020-06999. J.M. also acknowledges support from NSERC Discovery Grant No.~RGPAS-2020-00064, the CRC Program, a Government of Alberta MIF Grant, a Tri-Agency NFRF Grant (Exploration Stream), and the PIMS CRG program. D. Z. and Z.Y. are supported by the National Natural Science Foundation of China (Grants No.~11904417 and No.~
 12174455) and the Natural Science Foundation of Guangdong Province
 (Grant No.~2021B1515020026).
 \end{acknowledgments}
\bibliography{ref.bib}
\end{document}